\documentclass{article}

\PassOptionsToPackage{numbers, compress}{natbib}
\usepackage[preprint]{neurips_2022}

\usepackage{amsmath,amssymb,amsfonts}
\usepackage{graphicx}
\usepackage[utf8]{inputenc} % allow utf-8 input
\usepackage[T1]{fontenc}    % use 8-bit T1 fonts
\usepackage{hyperref}       % hyperlinks
\usepackage{url}            % simple URL typesetting
\usepackage{booktabs}       % professional-quality tables
\usepackage{amsfonts}       % blackboard math symbols
\usepackage{nicefrac}       % compact symbols for 1/2, etc.
\usepackage{microtype}      % microtypography
\usepackage{xcolor}         % colors

\title{Subject-specific quantitative susceptibility mapping using patch based deep image priors}

\author{%
 Arvind ~Balachandrasekaran, Davood ~Karimi, Camilo ~Jaimes, Ali ~Gholipour \\
 Computational Radiology lab\\
 Harvard Medical School\\
 Boston, MA 02115 \\
 \texttt{arvind.balachandrasekaran, davood.karimi@childrens.harvard.edu} \\
 \texttt{camilo.jaimescobos, ali.gholipour@childrens.harvard.edu} \\
}
\begin{document}
\maketitle
\begin{abstract}
Quantitative Susceptibility Mapping is a parametric imaging technique to estimate the magnetic susceptibilities of biological tissues from MRI phase measurements. This problem of estimating the susceptibility map is ill posed. Regularized recovery approaches exploiting signal properties such as smoothness and sparsity improve reconstructions, but suffer from over-smoothing artifacts. Deep learning approaches have shown great potential and generate maps with reduced artifacts. However, for reasonable reconstructions and network generalization, they require numerous training datasets resulting in increased data acquisition time. To overcome this issue, we proposed a subject-specific, patch-based, unsupervised learning algorithm to estimate the susceptibility map. We make the problem well-posed by exploiting the redundancies across the patches of the map using a deep convolutional neural network. We formulated the recovery of the susceptibility map as a regularized optimization problem and adopted an alternating minimization strategy to solve it. We tested the algorithm on a 3D invivo dataset and, qualitatively and quantitatively, demonstrated improved reconstructions over competing methods.

\end{abstract}
\section{Introduction}
Quantitative Susceptibility Mapping (QSM) is a parametric imaging technique to estimate the magnetic susceptibilities of biological tissues. It enables quantification of iron concentration in gray matter \citep{haacke2005imaging}, myelin content in white matter \citep{lee2012contribution}, deoxyhemoglobin in veins \citep{haacke1995vivo} etc. Changes in the concentrations reflect an underlying pathology, which include micro bleeds \citep{liu2012cerebral}, hemorrhages \citep{liu2013nonlinear} and neurodegenerative diseases \citep{lotfipour2012high,acosta2013vivo}, and can potentially be used as a bio-marker \citep{schenck2004high}. 

The susceptibility map is related to phase measurements through a 3D convolution with a dipole kernel. However, the problem of estimating the map through direct inversion is ill posed and results in streaking artifacts. To improve the quality of reconstructions, regularized recovery approaches enforcing signal sparsity and smoothness have been introduced \citep{bilgic2014fast,langkammer2015fast}. These approaches reduce the streaking artifacts but introduce blurring in the recovered maps. Recently, supervised deep learning methods have shown great potential and have generated maps with reduced artifacts \citep{jung2022overview}. However, to achieve reasonable reconstruction quality and for the network to generalize well to unseen data, these approaches require a lot of training data, which result in increased data acquisition time.

We propose a subject-specific, patch-based, unsupervised deep learning algorithm to estimate the susceptibility map from phase measurements. Patch based priors have been successful in many image restoration tasks including recovery of MR images from undersampled measurements \citep{ravishankar2010mr}. To make the problem well-posed, we exploit the redundancies across the different patches in the map using a deep convolutional neural network (CNN). The recovery of the map is then posed as a regularized optimization problem, where the representation of different patches by the deep neural network acts as a regularizer. We propose an alternating minimization strategy to solve the optimization problem and recover the susceptibility map. While the algorithm is developed for the QSM application, the proposed framework can be applied to solve any general linear inverse problem characterized by a forward model, of which denoising happens to be a special case \citep{asim2019patchdip}. We tested the algorithm on an invivo 3D Gradient echo dataset that was provided as part of a QSM reconstruction challenge \citep{langkammer2018quantitative}. We demonstrated improved QSM reconstructions, both quantitatively and qualitatively, over competing methods.

\section{Methods}

The forward model relating the magnetic resonance (MR) tissue phase $\boldsymbol{\Phi} \in \mathbb{R}^{N}$  and susceptibility map $\boldsymbol{\chi} \in \mathbb{R}^{N}$ is given by $\boldsymbol{\Phi} =\mathbf{F}^{-1}\mathbf{D}\mathbf{F}\boldsymbol{\chi} + \boldsymbol{\eta}$. N is the product of the spatial dimensions $(N_{x}, N_{y}, N_{z})$, $\mathbf{F}$ is the 3D Discrete Fourier Transform matrix, $\mathbf{D}=diag(\mathbf{d})$ where $\mathbf{d}[\mathbf{k}] = \frac{1}{3} - \frac{\mathbf{k}_{z}^{2}}{\mathbf{k}^2}, \mathbf{k} \neq 0$ is the spectrum of the dipole kernel, $\mathbf{k}^{2} = \mathbf{k}_{x}^{2}+\mathbf{k}_{y}^{2} + \mathbf{k}_{z}^{2}$ with $\mathbf{k}_{x}, \mathbf{k}_{y}, \mathbf{k}_{z}$ being the $\mathbf{k}$-space coordinates and $\boldsymbol{\eta}$ is the measurement noise. The problem of estimating $\boldsymbol{\chi}$ from $\boldsymbol{\Phi}$ is ill-posed. We propose to exploit the common features present in the patches of the susceptibility map and use it as prior information during recovery. Specifically, the redundancies across the patches of $\boldsymbol{\chi}$ are captured using a CNN. The recovery of $\boldsymbol{\chi}$ from $\boldsymbol{\Phi}$ is then formulated as the following regularized optimization problem:
\begin{equation}
\label{eq:problem formulation}
    {\boldsymbol{\Theta}^{*}, \boldsymbol{\chi}^{*}} = \arg \min_{\Theta, \boldsymbol{\chi}} \|\boldsymbol{\Phi} - \mathbf{F}^{-1}\mathbf{D}\mathbf{F}\boldsymbol{\chi}\|_{2}^2 + \mu \sum_{(i,j,k)}\|\mathbf{W}_{ijk}(\mathcal{R}_{ijk}(\boldsymbol{\chi}) - f_{\boldsymbol{\Theta}}(z_{ijk})\|^2_{2}
\end{equation}
In \ref{eq:problem formulation}, the first term is the data fidelity term, the second term acts as a regularizer and $\mu > 0$ is a regularization parameter. $\mathcal{R}_{ijk}(\boldsymbol{\chi}) \in \mathbb{R}^{p}$ extracts a patch of size ${p_{x} \times p_{y} \times p_{z}}$ from the location indexed by $(i,j,k)$ and vectorizes it. $f_{\boldsymbol{\Theta}}(.)$ is a CNN parameterized by the weights $\boldsymbol{\Theta}$ and $z_{ijk} \in \mathbb{R}^{p}$ is the input to the network. According to \citep{ulyanov2018deep}, a deep neural network with randomly initialized weights and random noise as input can be used to fit any degraded image under an observation model. Accordingly, we also assume $z_{ijk}$ to be uniform random noise with variance 0.1. $\mathbf{W}_{ijk}$ is a diagonal matrix where each entry (corresponding to a voxel location) is inversely related to the number of patches covering that voxel. Multiplication by the diagonal matrix is necessary to ensure that the same $\mu$ is applied to every voxel, which will not occur when a stride $> 1$ is used for patch extraction. We adopt an alternating minimization strategy to solve \ref{eq:problem formulation}, which decouples it into two simpler sub-problems corresponding to denoising and inversion steps. We iterate between the two steps until convergence is reached. $f_{\boldsymbol{\Theta}}(.)$ is chosen to be a 3D UNet architecture with four encoder and decoder levels with skip connections. In the denoising step, the network weights are initialized using the weights from the previous iteration. They are updated using an ADAM optimizer with a learning rate of 1e-6. We choose a patch size $=64 \times 64 \times 64$ and stride $=32 \times 32 \times 32$. The algorithm was run on a CentOS 7 Linux machine equipped with NVIDIA RTX A6000 GPU. 

We validated the proposed method on a dataset corresponding to a healthy subject, which was provided as part of the 2016 QSM reconstruction challenge \citep{langkammer2018quantitative}. The dataset $(160 \times 160 \times 160)$ was acquired on a 3T Siemens scanner with an isotropic resolution of 1.06 mm using a 3D Gradient echo sequence. QSM map computed from data acquired from multiple orientations (COSMOS) \citep{liu2009calculation} was used as ground truth. For comparison purposes, QSM maps were recovered using thresholded k-space division (TKD) \citep{shmueli2009magnetic} and regularized approaches using total variation (TV) and total generalized variation (TGV) penalties. The recovered maps corresponded to the lowest root mean square (rmse) with the ground truth.

\section{Results}

\begin{figure}
  \centering
  \includegraphics[trim=3cm 0.25cm 1cm 0.25cm,clip,width=1.16\linewidth]{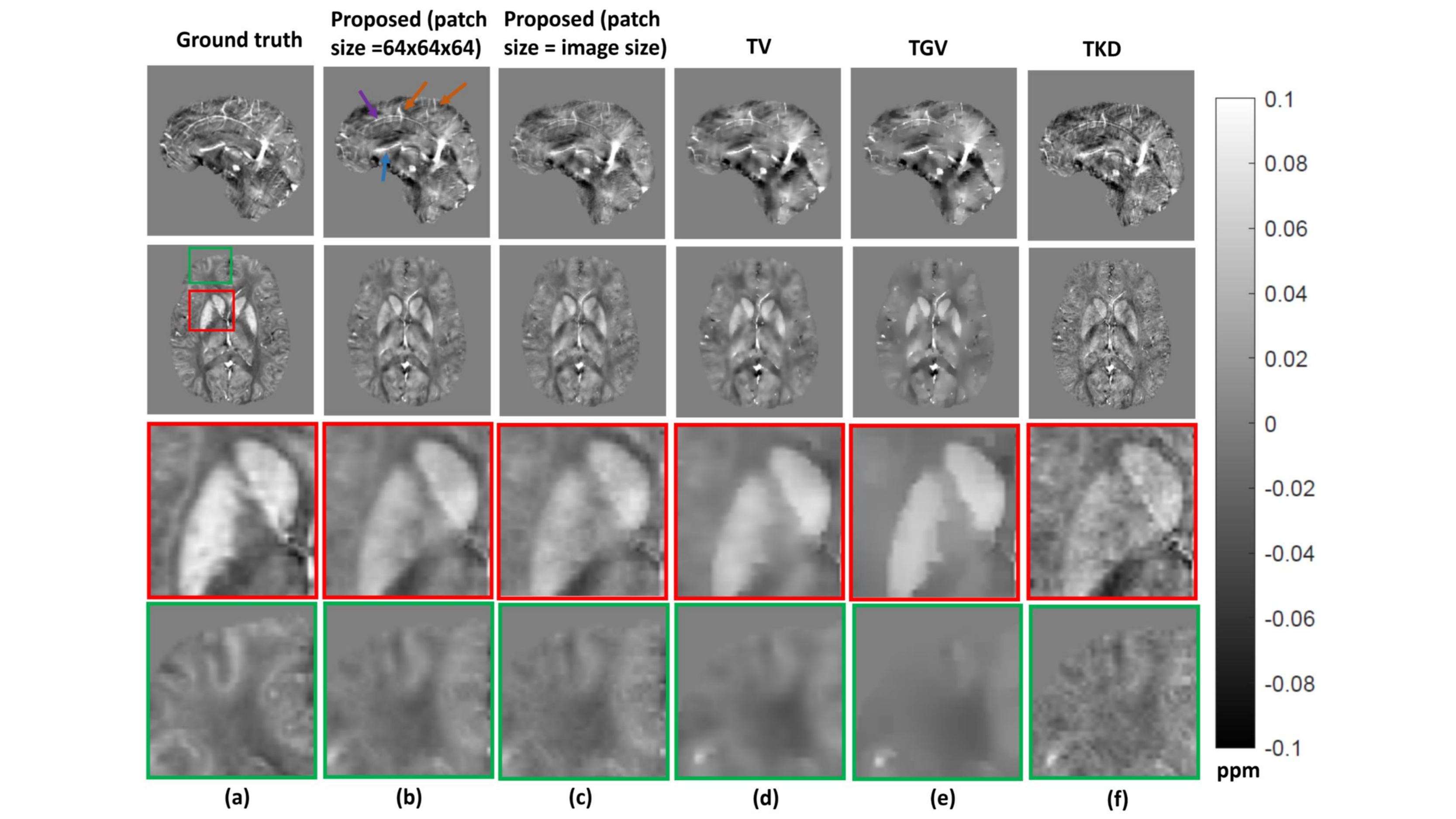}
  \caption{Comparison of the susceptibility maps (sagittal-1st row, axial-2nd row) recovered using different methods: (a) Ground truth, (b)  Proposed (patch size = $64\times64\times64)$, (c) Proposed (patch size = image size), (d) Total Variation (TV), (e) Total Generalized Variation (TGV), and (f) Thresholded k-space division (TKD).}
  \label{fig:1}
\end{figure}

The susceptibility maps corresponding to the different methods are shown in Fig. \ref{fig:1}\ref{fig:1}. Clearly, the proposed QSM reconstructions have better delineation of the finer structures such as the inferior sagittal sinus, cortical veins, and the deep veins, which are marked by purple, brown and blue arrows respectively. In addition, the proposed maps retain higher contrast between the gray and white matter structures, especially in the regions enclosed by the green and red boxes (rows 3 and 4 in Fig. \ref{fig:1}\ref{fig:1}). Also, the proposed maps are less noisy and more faithful to the ground truth when a smaller patch size is employed (see (b) vs (c)). This is because the weights of the network are being optimized for multiple smaller patches, which lead to the common features being learned while exhibiting high impedance towards learning noise. However, this comes at the cost of increased computation time. Quantitative metrics, shown in table \ref{tab:label 1}\ref{tab:label 1}, also indicate that the proposed method provides improved QSM reconstructions compared to the different methods.  

\begin{table}[!htpb]
  \caption{Comparison of various metrics for different methods}
  \label{tab:label 1}
  \centering
  \begin{tabular}{llll}
%    \toprule
%    \multicolumn{2}{c}{Part}                   \\
%    \cmidrule(r){1-2}
    Name     & RMSE     & SSIM  & PSNR  \\
    \midrule
    TKD & 70.57  & 0.79  & 40.55   \\
    \hline
    TGV & 61.4  & 0.83  & 41.76   \\
    \hline
    TV & 62.35  & 0.84  & 41.63   \\
    \hline
    Proposed (patch size = image size) & 59  & 0.85  & 42.1  \\
    \hline
    Proposed (patch size = 64x64x64) & \textbf{57.5}  & \textbf{0.86}  & \textbf{42.32} \\
    \bottomrule
  \end{tabular}
\end{table}

\section{Conclusion}
By proposing a subject-specific, unsupervised deep learning formulation for estimating the quantitative susceptibility map, we by-passed the generalization issue pertaining to current deep learning methods. We exploited the common features present in the patches of the map using a deep convolutional network and enforced it as a prior for the QSM inversion problem. We estimated the map by solving a regularized optimization problem using an alternating minimization strategy. As the weights of the network are optimized for multiple smaller patches, the network is more favorable to learn the redundancies across the patches, and will show high resistance to learning noise. Consequently, we observed that the proposed QSM reconstructions have better contrast between gray and white matter, delineation of finer venous structures, and is more faithful to the ground truth compared to the competing methods. Future work would involve extensive testing on a number of real datasets and validation of the performance through the scores of Neuroradiologists.

\section{ Potential Negative Societal Impact}
The authors do not identify any potential negative impact for this work on the society.

\bibliographystyle{unsrtnat}
\bibliography{ref}

\end{document}